\documentclass[aps,superscriptaddress,twocolumn]{revtex4-2}

\usepackage[utf8]{inputenc}
\usepackage[version=4]{mhchem}
\usepackage{textcomp}
\usepackage{graphicx}
\usepackage{float}
\usepackage{tabularx}
\usepackage{booktabs}
\usepackage{textcomp}
\usepackage{multirow}
\usepackage{placeins}

\newcommand{\degC}{\textdegree{}C}
\newcommand{\degph}{\textdegree{}C/h}
\newcommand{\TM}{\ce{TaTMTe4}}
\newcommand{\Ir}{\ce{TaIrTe4}}
\newcommand{\Rh}{\ce{TaRhTe4}}
\newcommand{\Ru}{\ce{TaRuTe4}}
\newcommand{\Rux}{\ce{Ta_{1+x}Ru_{1-x}Te4}}
\newcommand{\RuEDX}{\ce{Ta_{1.26(2)}Ru_{0.75(2)}Te_{4.000(8)}}}

\newcommand{\Mo}{\ce{MoTe2}}

\newcommand{\tMo}{T$_d$-\ce{MoTe2}}
\newcommand{\W}{\ce{WTe2}}
\newcommand{\IrRh}{\ce{TaIr_{1-x}Rh_{x}Te4}}
\newcommand{\RhIr}{\ce{TaRh_{1-x}Ir_{x}Te4}}

\begin{document}
\title{Layered van~der~Waals topological metals of \ce{TaTMTe4} (TM = Ir, Rh, Ru) family}

\author{G.~Shipunov}
\email{g.shipunov@ifw-dresden.de}
\affiliation{Institute for Solid State Research, Leibniz IFW Dresden, Helmholtzstr. 20, 01069 Dresden, Germany}
\author{B.R.~Piening}
\affiliation{Institute for Solid State Research, Leibniz IFW Dresden, Helmholtzstr. 20, 01069 Dresden, Germany}
\author{C.~Wuttke}
\affiliation{Institute for Solid State Research, Leibniz IFW Dresden, Helmholtzstr. 20, 01069 Dresden, Germany}
\author{T.~A.~Romanova}
\affiliation{P.N. Lebedev Physical Institute, Russian Academy of Sciences, 119991 Moscow, Russia}
\author{A.~V.~Sadakov}
\affiliation{P.N. Lebedev Physical Institute, Russian Academy of Sciences, 119991 Moscow, Russia}
\author{O.~A.~Sobolevskiy}
\affiliation{P.N. Lebedev Physical Institute, Russian Academy of Sciences, 119991 Moscow, Russia}
\author{E.~Yu.~Guzovsky}
\affiliation{P.N. Lebedev Physical Institute, Russian Academy of Sciences, 119991 Moscow, Russia}
\author{A.~S.~Usoltsev}
\affiliation{P.N. Lebedev Physical Institute, Russian Academy of Sciences, 119991 Moscow, Russia}
\author{V.~M.~Pudalov}
\affiliation{P.N. Lebedev Physical Institute, Russian Academy of Sciences, 119991 Moscow, Russia}
\author{D.~Efremov}
\affiliation{Institute for Solid State Research, Leibniz IFW Dresden, Helmholtzstr. 20, 01069 Dresden, Germany}
\author{S.~Subakti}
\affiliation{Institute for Solid State Research, Leibniz IFW Dresden, Helmholtzstr. 20, 01069 Dresden, Germany}
\author{D.~Wolf}
\affiliation{Institute for Solid State Research, Leibniz IFW Dresden, Helmholtzstr. 20, 01069 Dresden, Germany}
\author{A.~Lubk}
\affiliation{Institute for Solid State Research, Leibniz IFW Dresden, Helmholtzstr. 20, 01069 Dresden, Germany}
\author{B.~B\"uchner}
\affiliation{Institute for Solid State Research, Leibniz IFW Dresden, Helmholtzstr. 20, 01069 Dresden, Germany}
\affiliation{Institute of Solid State and Materials Physics, Technische Universit\"at Dresden, 01062 Dresden, Germany}
\author{S.~Aswartham}
\email{s.aswartham@ifw-dresden.de}
\affiliation{Institute for Solid State Research, Leibniz IFW Dresden, Helmholtzstr. 20, 01069 Dresden, Germany}

\date{\today{}}

\begin{abstract}
  Layered van~der~Waals materials of the family \ce{TaTMTe4} (TM=Ir,
  Rh, Ru) are showing very interesting electronic properties. Here we
  report the synthesis, crystal growth and structural characterization
  of \Ir{}, \Rh{}, \IrRh{} ($x = 0.06$; 0.14; 0.78; 0.92) and \Rux{}
  single crystals. For \Rux{} off-stoichiometry is shown. X-ray powder
  diffraction confirms that \Rh{} is isostructural to \Ir{}.
  We show that all these compounds are metallic with diamagnetic behavior.
  \RuEDX{} exhibits an upturn in the resistivity at low temperatures which is strongly field dependent.
  Below $T \approx 4$K we
  observed signatures of the superconductivity in the \IrRh{}
  compounds for $x = 0.92$. Magnetotransport measurements on all
  samples show weak magnetoresistance (MR) field dependence that is
  typically quadratic-in-field. However, for \IrRh{} with
  $x\approx 0.78$, the MR has a linear term dominating
  in low fields that indicates the presence of Dirac cones in the
  vicinity of the Fermi energy. For \Rh{} series the MR is almost
  isotropic. We have performed electronic structure calculations for isostructural \Ir{} and \Rh{} together with the projected total density of states. The main difference is appearance of the Rh-band close to the Fermi level.
\end{abstract}

\maketitle

\section{Introduction}

\begin{figure}
  \includegraphics[width=0.9\columnwidth]{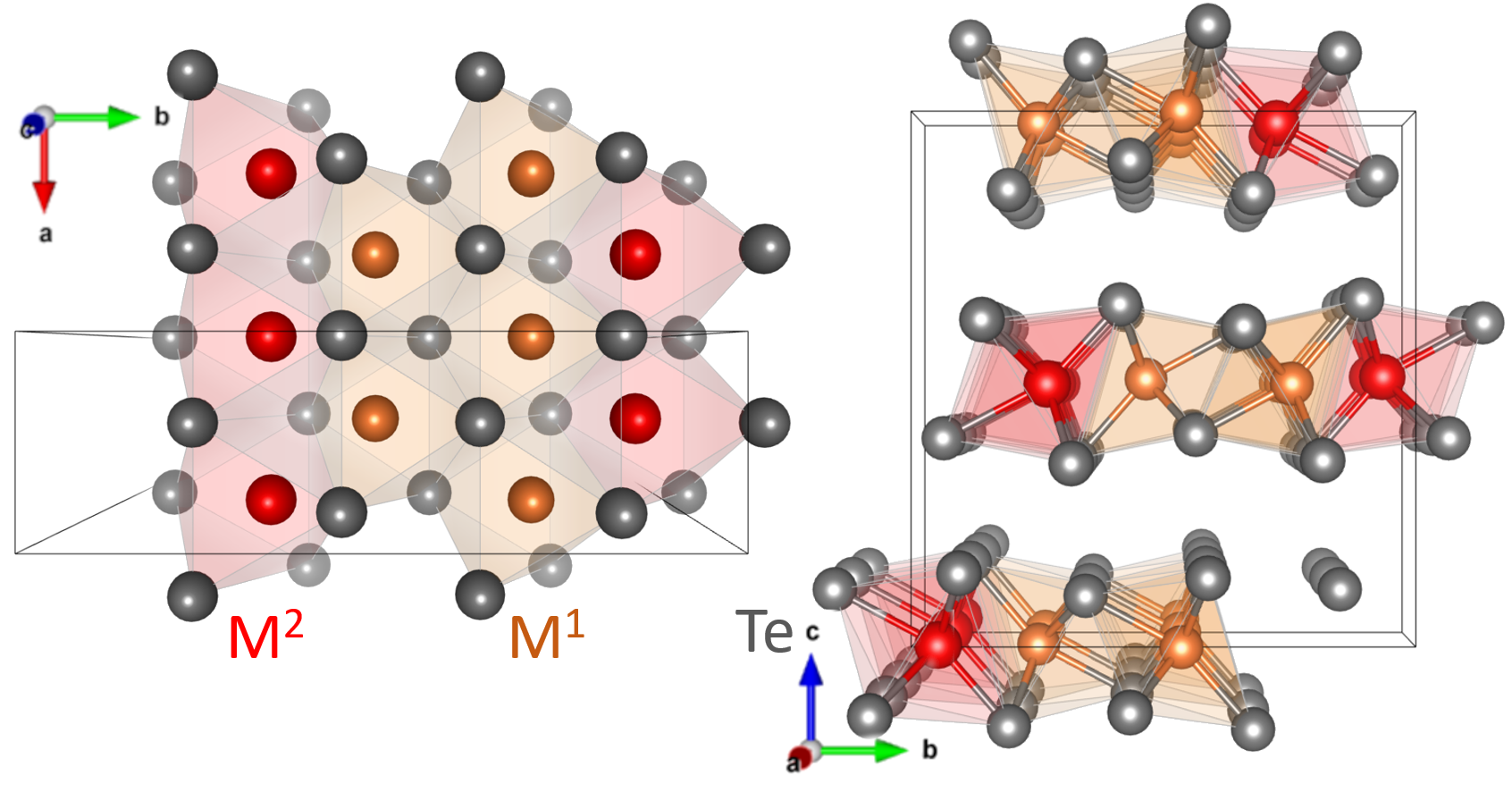}
  \caption{\label{fig:unitcell}
  Schematics of crystal structure of \ce{TMTe2} and \ce{TaTMTe4 compounds}. \ce{M^1}=\ce{M^2}=\ce{W} in case
of \W{} and \ce{M^1}=\ce{TM}, \ce{M^2}=\ce{Ta} for \TM{}. Note the reduced distance between \ce{M^1} and \ce{M^2} forming a zigzag chains along $a$.}
\end{figure}
Layered van~der~Waals two dimensional materials with sizable spin-orbit coupling effects such as \W{} and
\Mo{}, have been showing a wide array of fascinating
properties, which might be deliberately modified by a variety of parameters
, such as composition, thickness,
etc~\cite{manzeli2DTransitionMetal2017} makes them even more
attractive to gain insight into those physical effects. Recent reports
demonstrate type II Weyl semimetallic behavior, both theoretically and
experimentally, in both \tMo{} and \W{}
compound~\cite{liangElectronicEvidenceType2016,
  belopolskiFermiArcElectronic2016,
  huangSpectroscopicEvidenceType2016,
  wuObservationFermiArcs2016}. \ce{MoTe2} is also found to host an
edge supercurrent~\cite{wangEvidenceEdgeSupercurrent2020}, to be
superconductive~\cite{qiSuperconductivityWeylSemimetal2016} with
strong enhancement of $T_c$ in point-contact
measurements~\cite{naidyukSurfaceSuperconductivityWeyl2018} and with
sulfur
substitution~\cite{chenSuperconductivityEnhancementSdoped2016}
and in monolayers~\cite{Rhodes2021}.
Moreover,
\Mo{} and \W{} exhibit reversible metal-insulator transition,
making them attractive candidates for two-dimensional
nanoelectronics~\cite{zhangChargeMediatedReversible2016,
  naidyukSwitchableTopologicalDomains2020}.  \W{} shows similar array
of properties: the compound is known to host quantum spin Hall
state~\cite{tangQuantumSpinHall2017}, and has large non-saturating
magnetoresistance (MR)~\cite{aliLargeNonsaturatingMagnetoresistance2014,
  thoutamTemperatureDependentThreeDimensionalAnisotropy2015}.  Another
feature of this structure is the possibility of formation of polar
domains. As reported
in~\cite{sharmaRoomtemperatureFerroelectricSemimetal2019,
  huangPolarPhaseDomain2019}, \W{} and \Mo{} demonstrate switchable
spontaneous polarization and a natural ferroelectric domain structure,
\W{} even at room temperature. Combined with metallic
conductivity, this makes those materials first reported real-world
examples of a ferroelectric metal materials.

Due to all the fascinating
properties the search for materials with similar crystal structure is
an active area of research.
\Ir{} structure can be derived from \W{} by substituting all \ce{W} atoms with
\ce{Ta} and \ce{Ir} ordered into zigzag chains along $a$ direction,
where \ce{Ir} and \ce{Ta} atoms alternating along the
chain (Fig.~\ref{fig:unitcell})~\cite{marMetalmetalVsTelluriumtellurium1992,
  marSynthesisPhysicalProperties1992}. \Ir{} has been predicted to host Weyl
fermions~\cite{koepernikMathrmTaIrTeTernaryTypeII2016}, hosting only four
Weyl points (WPs), the minimal number of WPs a WSM with time-reversal
invariance can host. Later on signatures of
the WPs were found in
ARPES~\cite{hauboldExperimentalRealizationTypeII2017}.  Thickness-related
properties were also reported, e.g.\ in mono- and bilayer \Ir{}
quantum hall effect was reported~\cite{guoQuantumSpinHall2020}, while
bulk crystals show room-temperature nonlinear Hall effect with
temperature-induced sign
change~\cite{kumarRoomTemperatureNonlinear2020}. Moreover \Ir{} is
reported to host surface superconductivity with critical temperatures
of up to 1.54\,K~\cite{xingSurfaceSuperconductivityType2020a}. Among
other members of this ternary telluride family \ce{NbIrTe4} was
proposed to host type-II Weyl fermions with experimental indications
in charge transport
measurements~\cite{schonemannBulkFermiSurface2019}.  This material
also demonstrates non-saturating
MR~\cite{zhouNonsaturatingMagnetoresistanceNontrivial2019}. Recently
both binary (\W{}, \Mo{}) and ternary (\TM{}, TM=Ir, Rh, Ru)
tellurides showed bi-stable resistive metal-insulator switching in
point contact measurements at room temperature, bringing those
materials closer to
applications.~\cite{naidyukSwitchableTopologicalDomains2020}

Original reports of the \Ir{} structure also asserts the existence of several
other \ce{MM^{*}Te4} compounds and authors furthermore show that among other
compounds \Rh{} and \Ru{} exist and suggest that \Rh{} might
be isostructural to \Ir{}, while \Ru{} might be disordered analogue of
the \Ir{}.~\cite{marMetalmetalVsTelluriumtellurium1992,
  marSynthesisPhysicalProperties1992} This wide array of iso- or
similarly structured compounds makes this family a large playground
to study the dependence of physical properties on parameters such
as lattice size, interatomic distances, outer shell electron number,
etc.
Beyond initial structural studies of powder samples no further characterization
of structural and physical properties on single crystals have been reported.
Further chemical and structural investigation of those compounds
provides better description of the material at hand, as well as more
precise input parameters for theory models.

Here we report crystal growth of 3 members of \TM{} family, i.e. \Ir{}, \Rh{} and \RuEDX{}.
Also, we have successfully grown single crystals of doped series such as \IrRh{}.
Further, we discuss composition of acquired samples which
is followed by an assessment of the structure of the materials with
x-ray and electron diffraction methods. Finally we report results of
magnetotransport measurements for selected crystals.

\begin{figure*}
  \includegraphics[width=\textwidth]{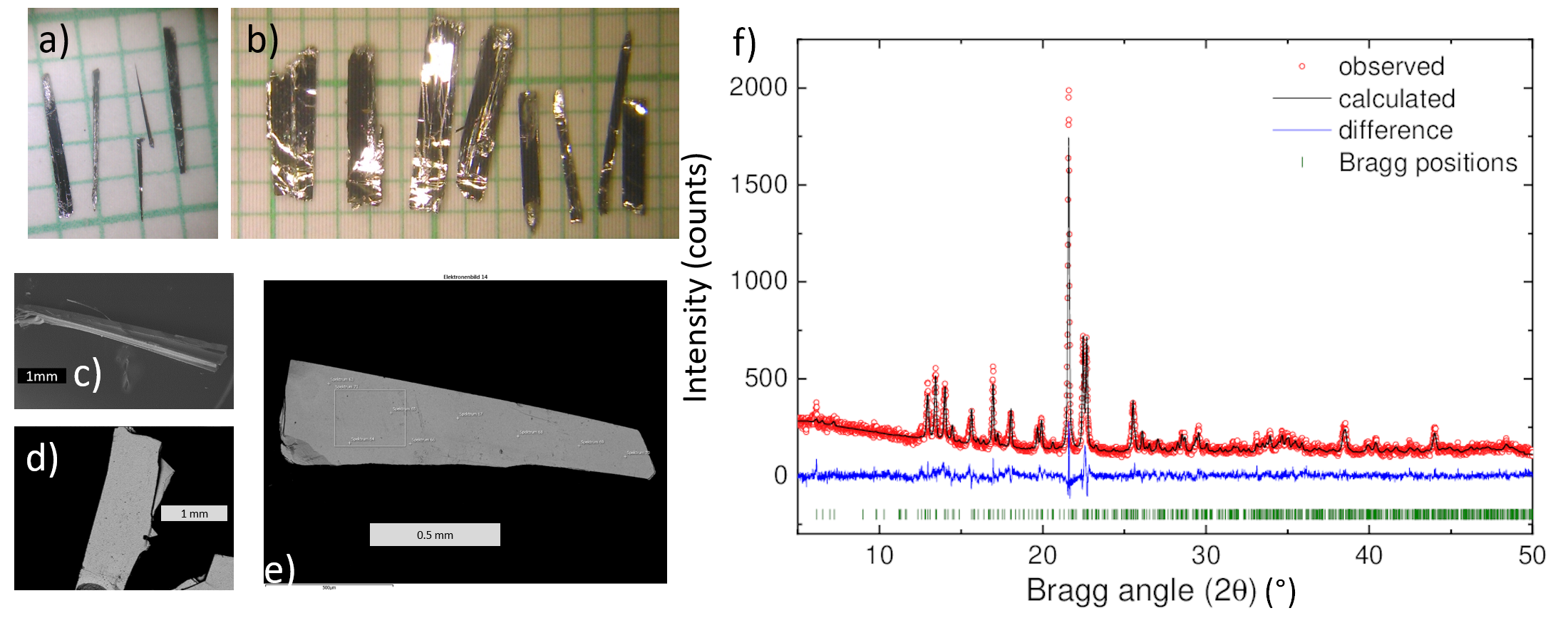}
  \caption{\label{fig:photos}\label{fig:SEM}\label{fig:rietveld:Rh}
    Images of as-grown \RuEDX{} (a) and \Rh{} (b) on a millimeter scale.
    SEM images of \Rh{} (c) and \RuEDX{} (d, e).
    X-ray powder diffraction plot with the fit by Rietveld method for \Rh{} ground crystals (f).
  }
\end{figure*}

\section{Experimental}
\subsection{crystal growth}

Single crystals of \Ir{}, \Rh{}, and \Rux{} were grown via self-flux
method.  Powders of \ce{Ta} (Alfa Aesar, powder 325 mesh, 99.97\%),
transition metal, \ce{TM=Ir} (Alfa Aesar, powder 325 mesh, 99.9\%),
\ce{Rh} (Evochem, powder, 99.95\%), \ce{Ru} (MaTecK, powder,
$<60\mu$m, 99.9\%) correspondingly, and \ce{Te} (Alfa Aesar, powder 18
mesh, 99.999\%) were mixed in \ce{Ta}:\ce{TM}:\ce{Te}=1:1:20 molar
ratio and thoroughly ground by hand in agate mortar.  Afterwards
reaction mixture was loaded into Canfield crucible
set~\cite{canfieldUseFritdiscCrucibles2016}, which in turn was placed
in evacuated fused quartz ampule.  The mixture was heated up
to 970\degC{} at 100\degph{}, dwelled at this temperature for 24\,h,
and subsequently slowly cooled at the rate of 1.5\degph{} to final
temperature of 600\degC{}.  Further, the crucible was taken out of hot
furnace, turned over and instantly put to a centrifuge which
facilitated flux removal.  Afterwards, crystal surfaces contaminated
by small amount of the flux were mechanically cleaved off before
further studies.  Furthermore, a substitution series of \IrRh{} was
synthesiszed. For growth of this substitution series mixture of elemental
powders with molar ratio \ce{Ta}:\ce{Ir_{1-x}Rh_x}:\ce{Te}=1:1:20
for $x=0.1$, 0.3, 0.7 and 0.9 was prepared as described for ternary
compounds, after which the mixture was heated to 1000\degC{}, held at
this temperature for 24\,h, and then cooled over course of 133\,h to
700\degC{}, after which the reaction mixture was centrifuged in
aforementioned manner.

\subsection{characterization}

\subsubsection{composition and structure}

The composition of the as grown single crystals was determined by
energy-dispersive x-ray spectroscopy (EDX), employing an electron beam probe in a scanning electron microscope
(accelerating voltage 30kV, current 552pA).  Structural characterization and
phase purity was confirmed by means of powder X-ray diffraction using a STOE
powder diffractometer (Bragg-Brentano transmission geometry, $2\theta$:$\omega$
scan, \ce{Co}~$K_{\alpha{}1}$ or \ce{Mo}~$K_{\alpha{}1}$ radiation, curved Ge
(111) monochromator).  Rietveld refinement of the x-ray data was carried out
with FullProf~\cite{fullprof} and Jana2006~\cite{jana} software packages.

Selected area electron diffraction (SAED) on thin exfoliated \TM{}
(TM=Ir, Ru, Rh) was utilized by employing FEI Tecnai F20 transmission
electron microscope (TEM) with \ce{LaB6} emitter
operated at 200~kV acceleration voltage. The \TM{} crystals used in
this work were mechanically exfoliated using commercially available
adhesive tape (eco, tesa). The exfoliated \TM{} crystals were
separated from the tape by immersion in 10~mL of acetone and
isopropanol (1:1) solution in 12~mL glass sample vial.
Ultrasonification (frequency $\approx{}35$~kHz) was employed to assist the
separation of \TM{}-flakes from the tape surface for two hours
($8 \times{} 15$~minutes sonication process with 5~minutes breaks
between the 15~minutes session).
The micrometer-sized \TM{} flakes were then transferred to the TEM lacey-carbon
Cu grids using a standard pipette. To ensure the cleanliness of the
\TM{}-flakes the elemental composition of every \TM{}-flake was
confirmed by in-situ TEM-EDX before collection of the SAED pattern.
Theoretical kinematic electron diffraction
patterns were computed and visualized using the SingleCrystal software
package version 3.1.5, (CrystalMaker Software LtD., UK).

\subsubsection{physical properties}

Magnetic susceptibility was measured using a MPMS superconducting
quantum interference device (SQUID) magnetometer from Quantum Design.
Transport and magnetotransport measurements with \ce{TaTMTe4}
compounds were performed in fields up to 9\,T using a PPMS-9 system from
Quantum Design, and up to 16\,T using CFMS-16 system from
Cryogenic respectively.
Temperature dependencies of the longitudinal resistivity
$\rho_{xx}$ were measured in the temperature range 2--300\,K, and
angular dependencies of $\rho_{xx}(H)$ (the magnetoresistance, MR)
at temperatures near 2\,K. The data on the type and density of
carriers were obtained from Hall effect measurements. Resistivity
values were calculated from the measured resistance and the
sample dimensions.  For the magnetotransport experiment we used the
``Hall bar'' geometry, where the current was always applied perpendicular to
the magnetic field direction.

\section{Results and Discussion}

\subsection{composition and structure}

The grown crystals depicted on fig.~\ref{fig:photos}~(a, b) are easily cleavable silvery flat needles of up to
0.5\,cm in length and up to 0.1\,cm in diameter with mass on the order
of tens of milligrams.  Crystals of \Ir{} show a needle-like
appearance, while \Rh{} and \RuEDX{} are more flattened, in a shape of
thin stripes.  As-grown crystals are extremely ductile and cleavable,
even under regular handling operations.
The results of the EDX confirm the nominal stoichiometry for \Rh{} and
\Ir{} samples.  Backscattered-electron SEM images (fig.~\ref{fig:SEM}~(c--e)) show uniform
intensity, which is indicative of the absence of macroscopic
inhomogeneities.
Interestingly, the nominal ratio Ru:Ta=1:1 as-grown crystals in fact show \RuEDX{}
composition.  This Ru:Ta ratio is consistent across different
measurement points and different crystals, on natural and freshly cleaved surfaces, with a low EDX
statistical error.  Assuming the same connectivity in the structure as
in \Ir{} compound, this composition shift must manifest as structural modification
hinted in~\cite{marMetalmetalVsTelluriumtellurium1992}.
This composition shift might be explained by the fact that Ru and Ta radii are quite close,
resulting in e.g. mixing of the atoms in the same crystallographic position.

Results of EDX measurements on the \IrRh{} are presented in
table~\ref{tab:EDX:IrRh}. We see \ce{Rh}/\ce{Ir}
substitution, with composition consistently shifted
towards metal with the higher concentration.

For characterization of the structure powder x-ray
diffraction (PXRD) was carried out including further analysis by Rietveld method.
Results of the x-ray powder diffraction on \Rh{} powdered crystals
are presented in fig.~\ref{fig:rietveld:Rh}~(f);
PXRD patterns for \RuEDX{} and \IrRh{} are presented on fig.~\ref{pxrd-comp} in appendix,
due to strong texture in both cases and high ductility in case of \RuEDX{} Rietveld analysis of those
didn't prove to be feasible.
One observes fig.~\ref{fig:rietveld:Rh} diffraction shows broad reflections, which
can be attributed to high ductility and ease of cleavage of the
material.  \Ir{} was refined in the structural model presented in ICSD~\cite{marMetalmetalVsTelluriumtellurium1992},
(ICSD \textnumero{}73322, space group $Pmn2_1$).  As reported by Mar
et al.~\cite{marMetalmetalVsTelluriumtellurium1992}, we see that in
case of \Rh{} compound diffraction pattern is quite similar to \Ir{}.
The \Rh{} diffraction pattern was clearly indexed in $Pmn2_1$ space group, so for
further Rietveld analysis we selected \Ir{} as initial structure model, with
\ce{Ir} positions substituted by \ce{Rh} with approximate lattice
parameters extracted in the indexing step.
Refinement in this model
yielded lattice parameters of $a=3.75670(11)$, $b=12.5476(5)$,
$c=13.166(3)$ and cell volume of $V=620.20(15)$ for \Rh{}.
We observe small change of $a$ and $c$ parameters
$\Delta a \approx -0.03$\AA; $\Delta c\approx -0.03$\AA, and a considerable increase of $b$ ($\Delta b\approx 0.12$\AA) compared to
the \Ir{} structure.
This can be explained by the smaller radius of the
\ce{Rh} atom, and as a result, might be indicative of tighter, and
more extended zigzag chains in the structure.  Refined atomic
positions are presented in table~\ref{wyckoff}.  With that we can
conclude that the \Rh{} compound is isostructural to \Ir{}, which provides
the opportunity to study change of physical properties with the
change of lattice parameters and isovalent substitution.

\begin{figure}
  \includegraphics[width=\columnwidth{}]{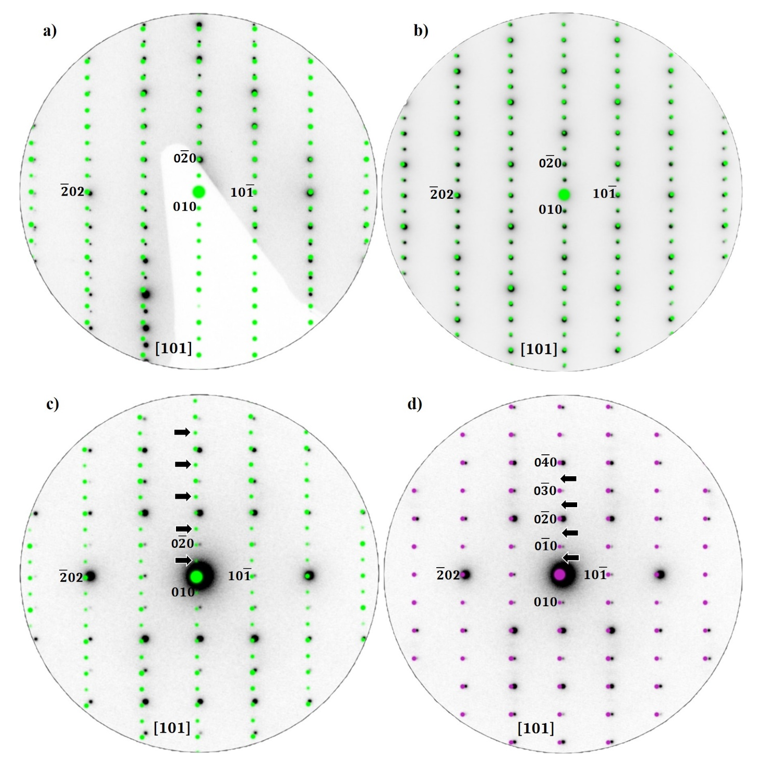}
  \caption{\label{fig:TEM}
    Selected area electron diffraction (SAED)
    on \Ir{}, \Rh{} and \RuEDX{}.
    (a,b) SAED patterns of
    representative \Ir{} and \Rh{} nanoflake oriented along [101] zone
    axis overlaid with theoretical SAED patterns (green dots) using
    $Pmn2_1$ (\textnumero{}31) space group obtained from PXRD data
    refinement results in this work. (c,d) SAED pattern of a
    representative \RuEDX{} nanoflake oriented along [101] zone axis
    overlaid with theoretical SAED patterns (green dots) using $Pmn2_1$
    (\textnumero{}31) structure derived from \Ir{} (c) and a $Pmn2_1$ structure derived from \W{}
    (unit cell is halved along $b$ compared to \Ir{}, d).
    The
    simulated reflections are intentionally slightly shifted for
    better overview. The arrows on the (c) and (d) images are given to
    mark the different fit between the two crystal models for \RuEDX{}.  }
\end{figure}

SAED on individual \TM{}
nanoflakes exfoliated from single crystals was performed to elaborate
the crystallographic information obtained from the PXRD data reported
in this work (Fig.~\ref{fig:TEM}). The SAED patterns for \TM{}
acquired in [101] orientation are compared with simulated electron
diffraction (ED) pattern using $Pmn2_1$ (\textnumero{}31) space group.
Here the
[101] zone axis allows to distinguish lattice modulations in ternary \TM{}
systems along $b$-direction in
reciprocal space.
The good agreement of all symmetrical equivalent reflections
in the ED patterns for \Ir{} and \Rh{}
strongly suggest that these two crystals are isostructural and
crystallize in $Pmn2_1$ space group as described in
\cite{marMetalmetalVsTelluriumtellurium1992}.
Interestingly, the \RuEDX{} electron diffraction patterns show
that the unit cell of this compound is halved along $b$ compared to \Ir{}.
To obtain a better fit with the SAED pattern we therefore used a crustal structure for \RuEDX{},
which was derived from \W{} phase by filling the two \ce{W} 2a Wyckoff sites with \ce{Ta}
and \ce{Ru} respectively.
The proposed new structure matches nicely to the
experimental SAED data of \RuEDX{} compound (purple dots,
Fig.~\ref{fig:TEM}~d) and hence also provides a hint concerning the
anomaly related to weak $k = 2n + 1$ reflections reported in
\cite{marMetalmetalVsTelluriumtellurium1992}.
Note, however, that this new structure does not reflect the large off-stoichiometry and the atomic positions and exact symmetry remain unclear as of now.
Further studies are therefore required to fully resolve the \RuEDX{} structure.

\subsection{physical properties}

\begin{figure}
    \includegraphics[width=\columnwidth]{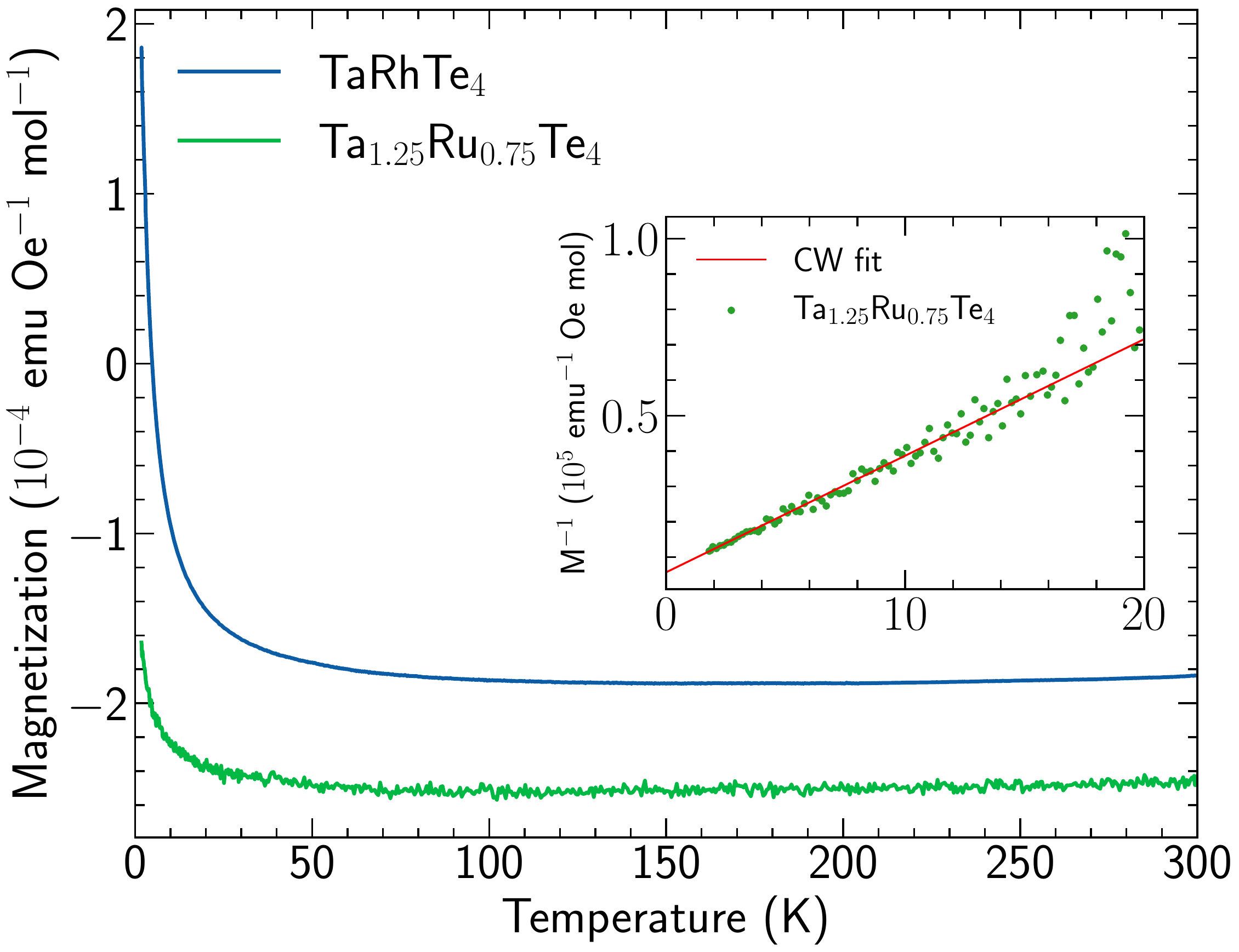}
    \caption{\label{fig:squid:chi}
      Temperature dependence of molar magnetization for \Rh{} and \RuEDX{}; Inset: Curie-Weiss fit of paramagnetic
      contribution to \RuEDX{} magnetization.
    }
\end{figure}

\begin{figure}
  \includegraphics[width=\columnwidth]{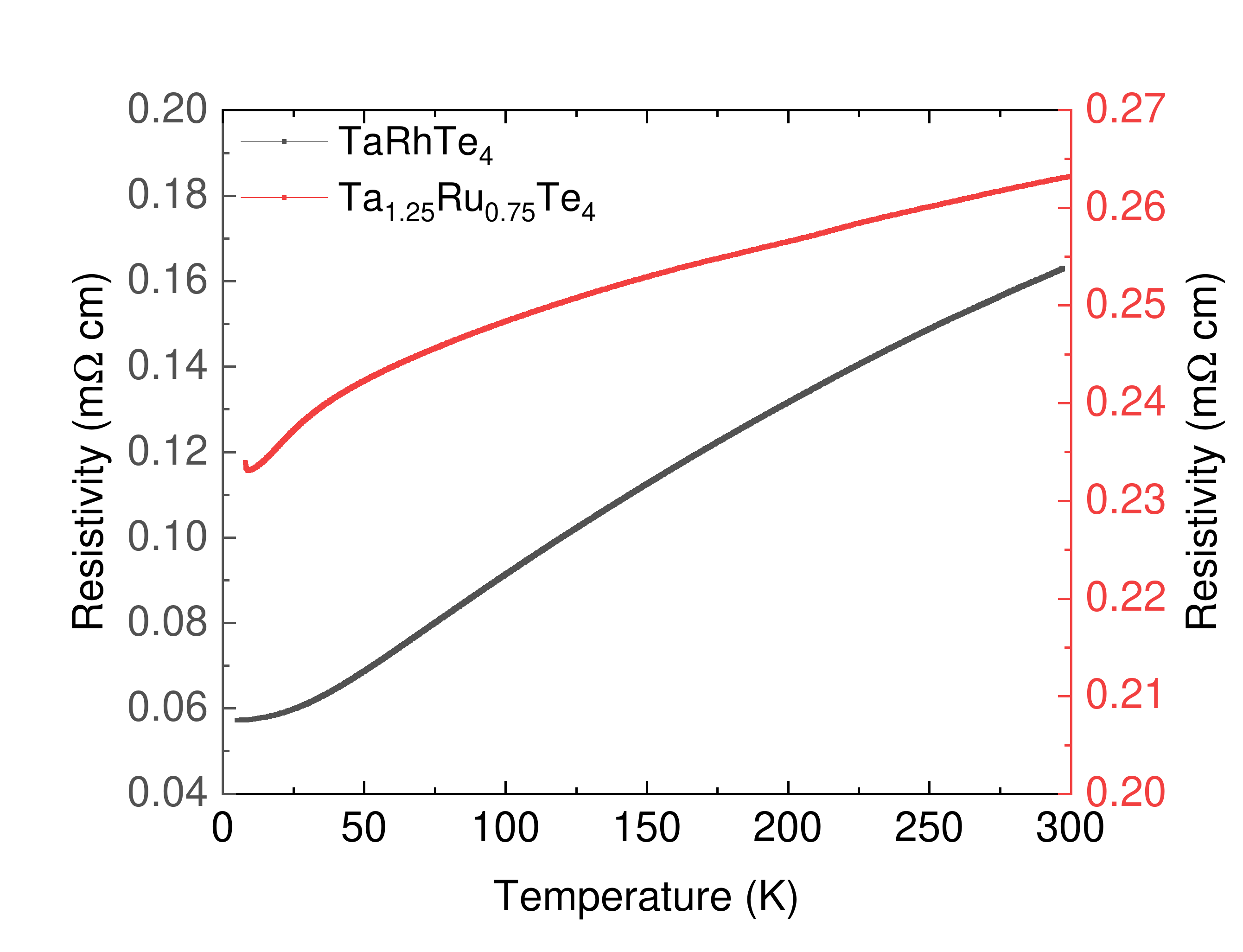}
    \caption{\label{fig:res} Temperature dependence of the
    resistivity for \RuEDX{} and \Rh{}. }
\end{figure}

\begin{figure*}
  \includegraphics[height=6cm]{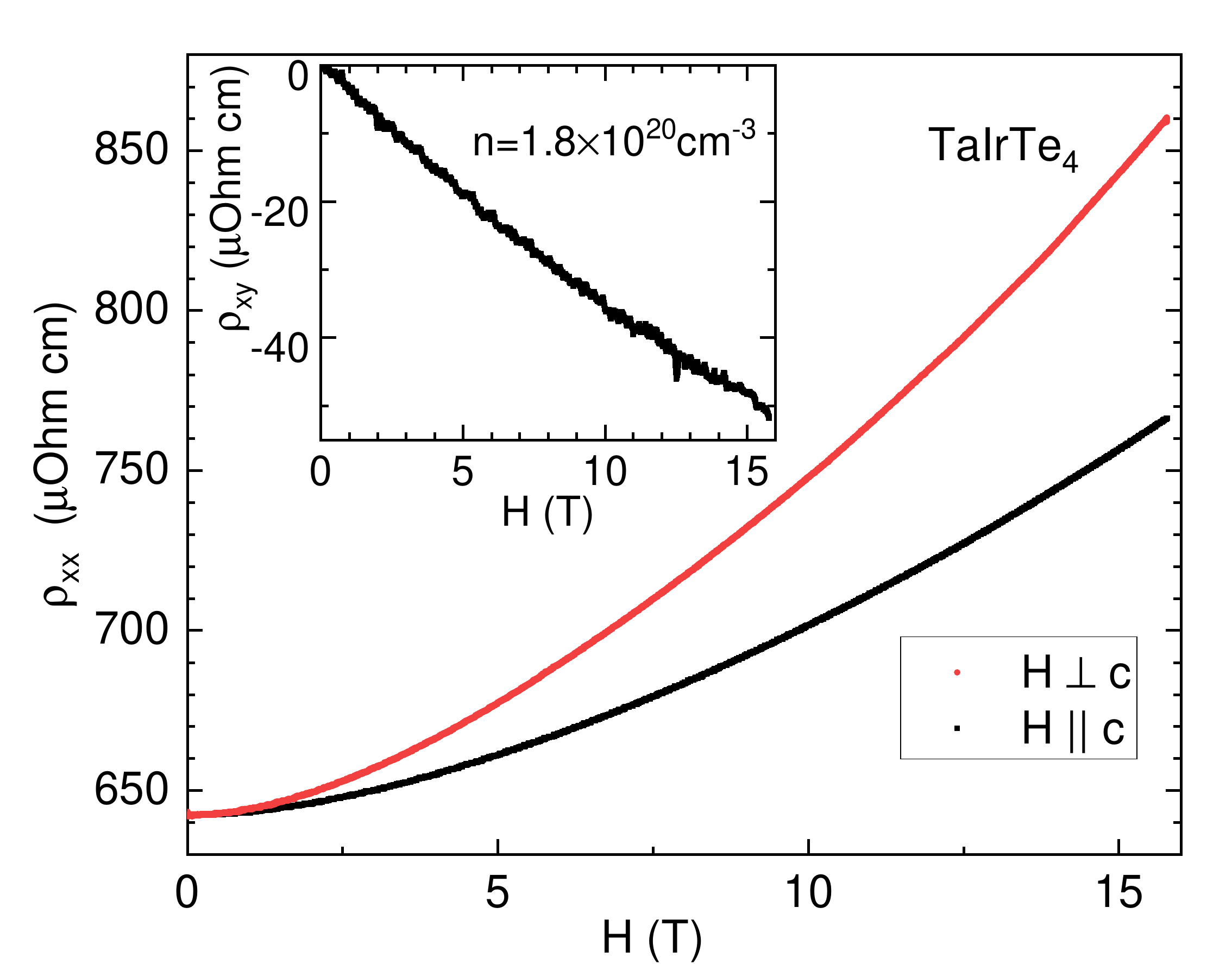}
  \includegraphics[height=6cm]{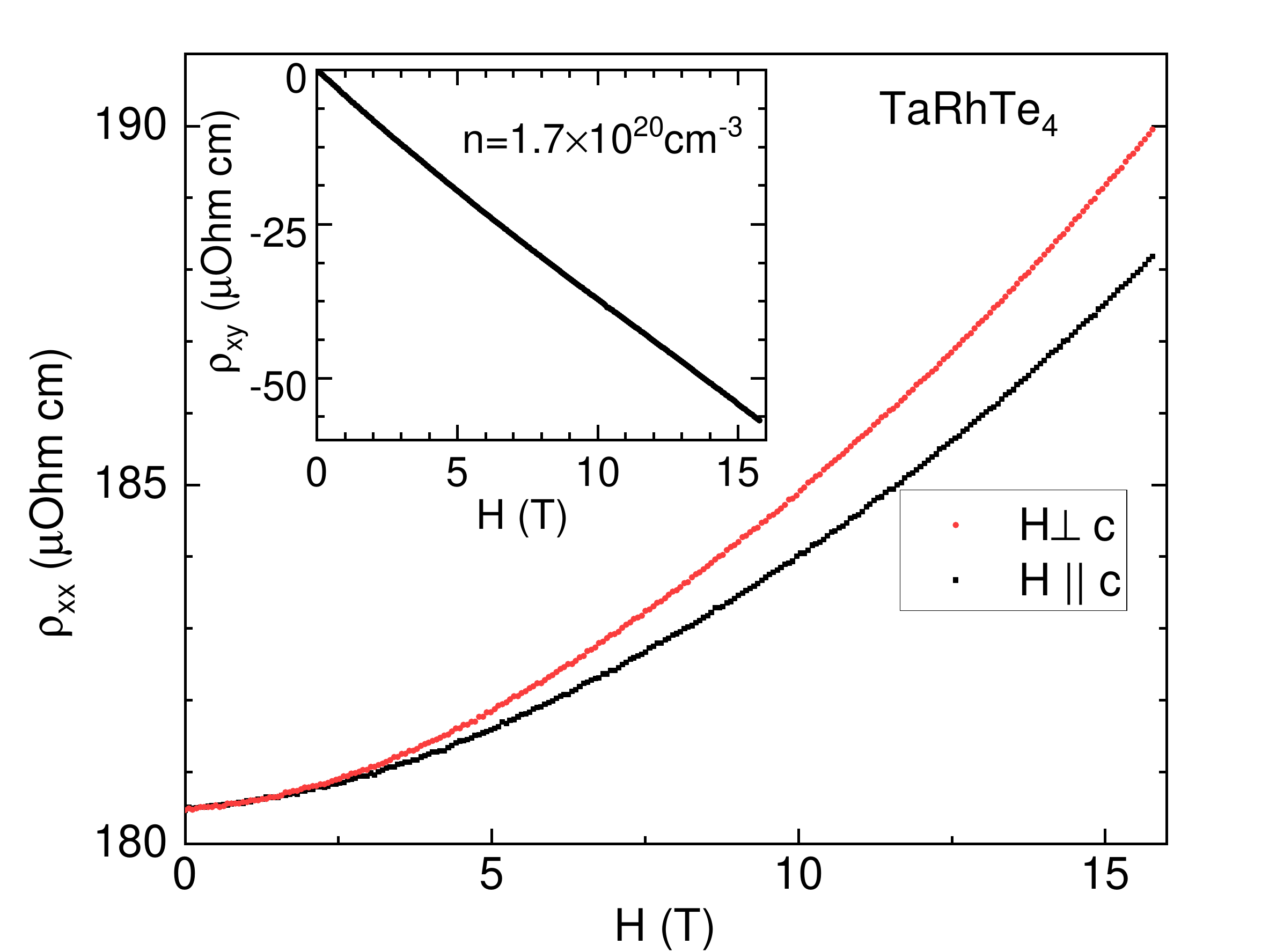}
  \caption { Left panel: magnetic field dependence of the diagonal
    resistivity $\rho_{xx}$ for TaIrTe$_4$ at two field
    orientations. Inset: Hall resistivity $\rho_{xy}$ versus magnetic
    field. Temperature $T=2$\,K.  Right panel: magnetic field
    dependence of $\rho_{xx}$ for TaRhTe$_4$ at two field
    orientations. Inset: $\rho_{xy}$ versus magnetic
    field. $T=2K$. $n=1.7\times 10^{20}$cm$^{-2}$, as determined from
    Hall resistivity.}
  \label{fig:Rh}\label{fig:Ir}
\end{figure*}

\subsubsection{Magnetization}

For ternary compounds the temperature dependence of magnetization was
measured.
The derived temperature dependence of molar magnetization for \Rh{} and \RuEDX{} is presented in fig.~\ref{fig:squid:chi}.
All compounds exhibit diamagnetic behavior,
with Curie-Weiss tails at low temperatures.
Inset in Fig.~\ref{fig:squid:chi} shows the Curie-Weiss analysis for \RuEDX{} at temperatures below 20\,K.
Data confirms absence of any long range magnetic order down to 1.8\,K.

\subsubsection{Transport and magnetotransport}

For stoichiometric TaRhTe$_4$ (Fig.~\ref{fig:res}) as well as for
Ir-doped \RhIr{} compound with $x=0.08$ (Fig.~\ref{fig:RhIr008-SC}) the temperature
dependency of the diagonal resistivity in zero field flatten below
$\approx 20$\,K.  In magnetic field, both for $B\parallel b$ and
$B\parallel c$ (see Fig.~\ref{fig:Rh}), the resistivity also follows a
classical almost isotropic quadratic dependence.  The bulk carrier
density determined from Hall resistance measurements (inset to fig.~\ref{fig:Rh} $n\approx 1.7\times 10^{20}$cm$^{-3}$ for the
stoichiometric TaRhTe$_4$ is by a factor of two lower compared to well investigated member of the family TaIrTe$_4$.
As shown in Fig.~\ref{fig:Rh}, the resistivity of \Ir{} varies quadratically
with field which is quite similar in our case of \Rh{}.
\Ir{} resistivity exhibits a minor (8\% at $B=9$\,T) anisotropy.
The Hall resistivity $\rho_{xy}$ is negative and
linearly grows with field, thus indicating dominant $n$-type electronic
states with a similar bulk carrier density of $1.8\times 10^{20}$cm$^{-2}$.
The initial minor
anisotropy of the magnetoresistance becomes even lower upon doping with
Rh or Ru.

\RuEDX{} demonstrate metallic-type transport behavior (Fig.~\ref{fig:res}) at
least down to 10\,K.
Although the resistivity shows a markedly different behavior at low
temperatures, where it exhibits a minimum at $T \approx 9$\,K, which appears to
be a kink followed by a steep upturn while further lowering the temperature
below 9\,K.
The origin of such non-monotonic resistivity remains unclear and requires further investigation.

The 22\% substitution of Rh by Ir does not change significantly the
resistivity $T$- dependence: RRR remains almost unchanged, whereas the
carrier density increases by 34\% to $2.3 \times 10^{20}$cm$^{-3}$,
and the magnetoresistance becomes fully isotropic.
Interestingly, for the intermediate Ir-doping of 8\%, the sample shows
signatures of the emerging superconducting (SC) transition (see
Fig.~\ref{fig:RhIr008-SC} in Appendix).

\subsubsection{Electronic structure calculations}
\begin{figure}
	\centering
	\includegraphics[width=\linewidth]{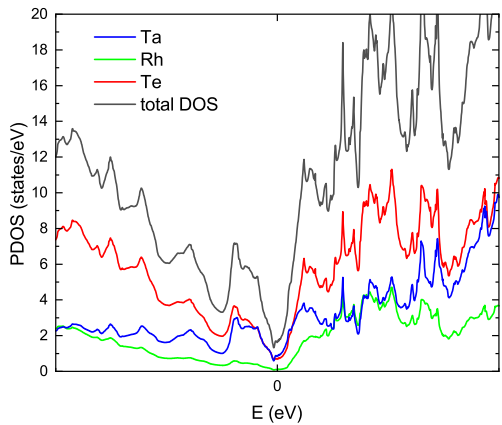}
    \caption{Total DOS of TaRhTe$_4$ and the partial contribution of Ta, Rh and Te orbitals. \label{fig.DOS}}
\end{figure}

\begin{figure*}
	\centering
	\includegraphics[width=\textwidth]{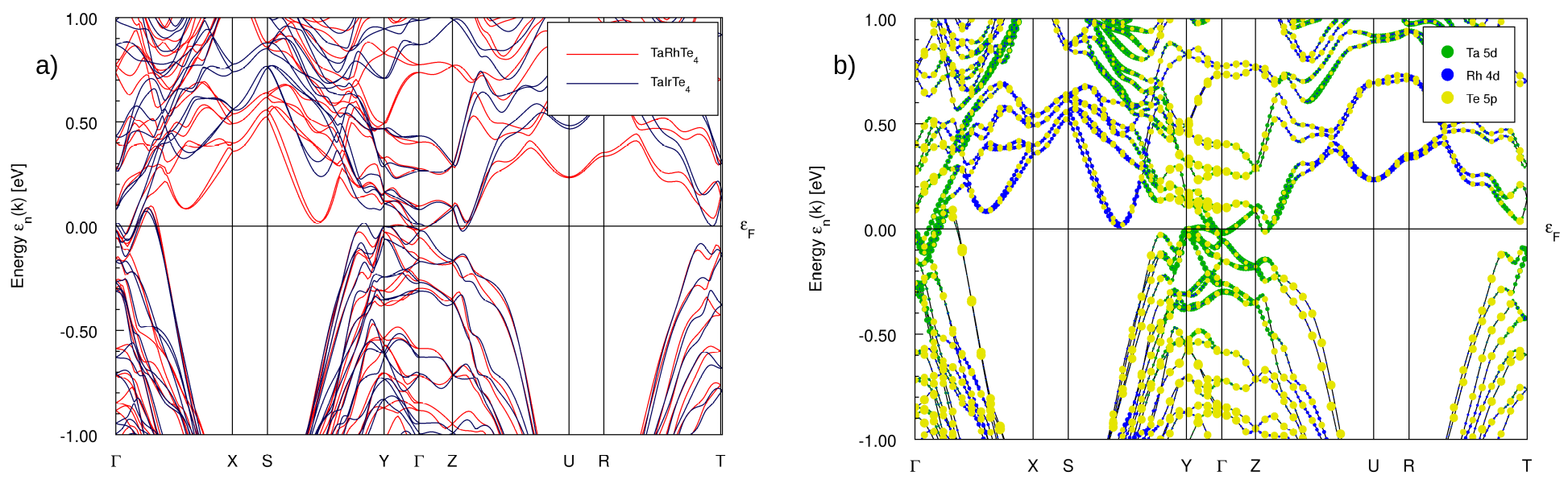}
    \caption{Band structure TaRhTe$_4$ vs TaIrTe$_4$ (a). Orbital contribution to the electronic band structure of TaRhTe$_4$ (b).
	\label{fig:fatbands2}
	\label{fig.bandstructure1}
    }
\end{figure*}

We have performed calculations on the electronic structure within the DFT theory.
The full relativistic generalized gradient approximation (GGA) in the
Perdew-Burke-Ernzerhof variant is used for the exchange correlation potential
implemented in the full potential local orbital band-structure package (FPLO)~\cite{FPLO,FPLO2}.
For  the  Brillouin zone (BZ) integration we used the tetrahedron method with $12\times12\times12 k$-mesh.
The obtained electronic density of states (DOS) is presented in Fig.~\ref{fig.DOS}  decreases slightly at the Fermi level and then increases continuously  similar TaIrTe$_4$.
Band structures of TaRhTe$_4$ with TaIrTe$_4$ are presented in Fig.~\ref{fig.bandstructure1}.
The main difference is appearance of the Rh-band close to the Fermi level.
A detailed look at electronic structure properties of \Rh{} together with investigation
via ARPES will be reported elsewhere~\cite{rh-inprep}.
The orbital contribution to the electronic band structure is given in the Fig.~\ref{fig:fatbands2}.
The rest of the bands are similar for both structures.
It explains the similarity in the physical properties of both materials.


\section{Conclusion}

In conclusion, we reported single crystal growth of layered van~der~Waals materials of the family \TM{}
(TM=Ir,Rh,Ru), as well as several \IrRh{} compounds.
We confirmed the
nominal compositions via SEM-EDX. In the \Rux{} case we demonstrate that the compound
is not stoichiometric with an actual composition of \RuEDX{}.
X-ray
powder diffraction and further Rietveld refinement confirms that \Rh{}
is isostructural to \Ir{} and crystallizes in non-centrosymmetric
orthorhombic $Pmn2_1$ space group with a similar connectivity as \ce{WTe2}.
  SAED
results indicate that \RuEDX{} is a disordered analog of \Ir{}
structure type and further investigation of the structure is needed.
Substitution series of \IrRh{} were
obtained for x = 0.06; 0.14, 0.78, 0.92 and we investigated magnetic
and magnetotransport properties of the grown single crystals.

All samples have electronic ($n$-type) conduction.  Any substitution of \ce{Rh} or \ce{Ir}
was found to reduce mobility.
All samples show weak MR field dependence.
For the \ce{Rh} series the MR is almost isotropic.
Some compounds show a linear MR in low fields, atop of the parabolic one.
\RuEDX{} shows parabolic magnetoresistance which flattens out in high fields at the temperatures above 9\,K.
Magnetoresistance measurements below 9\,K show a clear anomaly in fields below 5\,T.
We have performed electronic structure calculations using DFT for isostructural \Ir{} and \Rh{}
together with the projected total density of states.
The main difference is appearance of the Rh-band close to the Fermi level.

Due to the topological properties of the electronic band structure,
the recent observation of the resistive switching at the room
temperature, together with the observed signatures of the
superconducting transition,
these systems are very interesting for further investigation.

\paragraph*{\textbf{Acknowledgments}}

SA acknowledges financial support of Deutsche Forschungsgemeinschaft
(DFG) through Grant \textnumero{}AS 523/4-1.
SA, DE, BB acknowledge DFG financial support through DFG-RSF project \textnumero{}449494427.
BB acknowledges
financial support from the projects of the Collaborative Research
Center SFB 1143 at the TU Dresden (project-id 247310070) and
W\"urzburg-Dresden Cluster of Excellence on Complexity and Topology in
Quantum Matter ct.qmat (EXC 2147, project-id 390858490).  AVS, ASU and
VMP acknowledge support from RFBR (project \# 21--52--12043).
SS, DW and AL acknowledge funding from DFG SFB~1415, Project ID \textnumero{}417590517.
Transport and magnetotransport measurements were done using equipment of the LPI Shared Facility Center.
We acknowledge U.~Nitzsche for technical assistance.

\paragraph*{\textbf{Author contribution}}
Single crystal growth and characterization experiments, magnetization
measurements were performed by GS, BRP, BB and SA\@.
The resistivity and magnetoresistance measurements on \RuEDX{} were preformed
and analyzed by CW\@.
Resistivity and magnetoresistance measurements were performed and analyzed by
TAR, AVS, OAS, EYuG, ASU and VMP\@.
Selected area electron diffraction experiments were preformed and interpreted
by SS, DW and AL\@.
Electronic band structure calculations were preformed by DE.
GS and SA wrote the manuscript with input from all co-authors.

\paragraph*{\textbf{Data availability}}
The datasets generated during and/or analyzed during the current study
are available from the corresponding authors on reasonable request.

\paragraph*{\textbf{Competing interests}}
The authors declare no competing financial or non-financial interest.

\section{Appendix}

\begin{figure*}
\includegraphics[width=1.1\columnwidth{}]{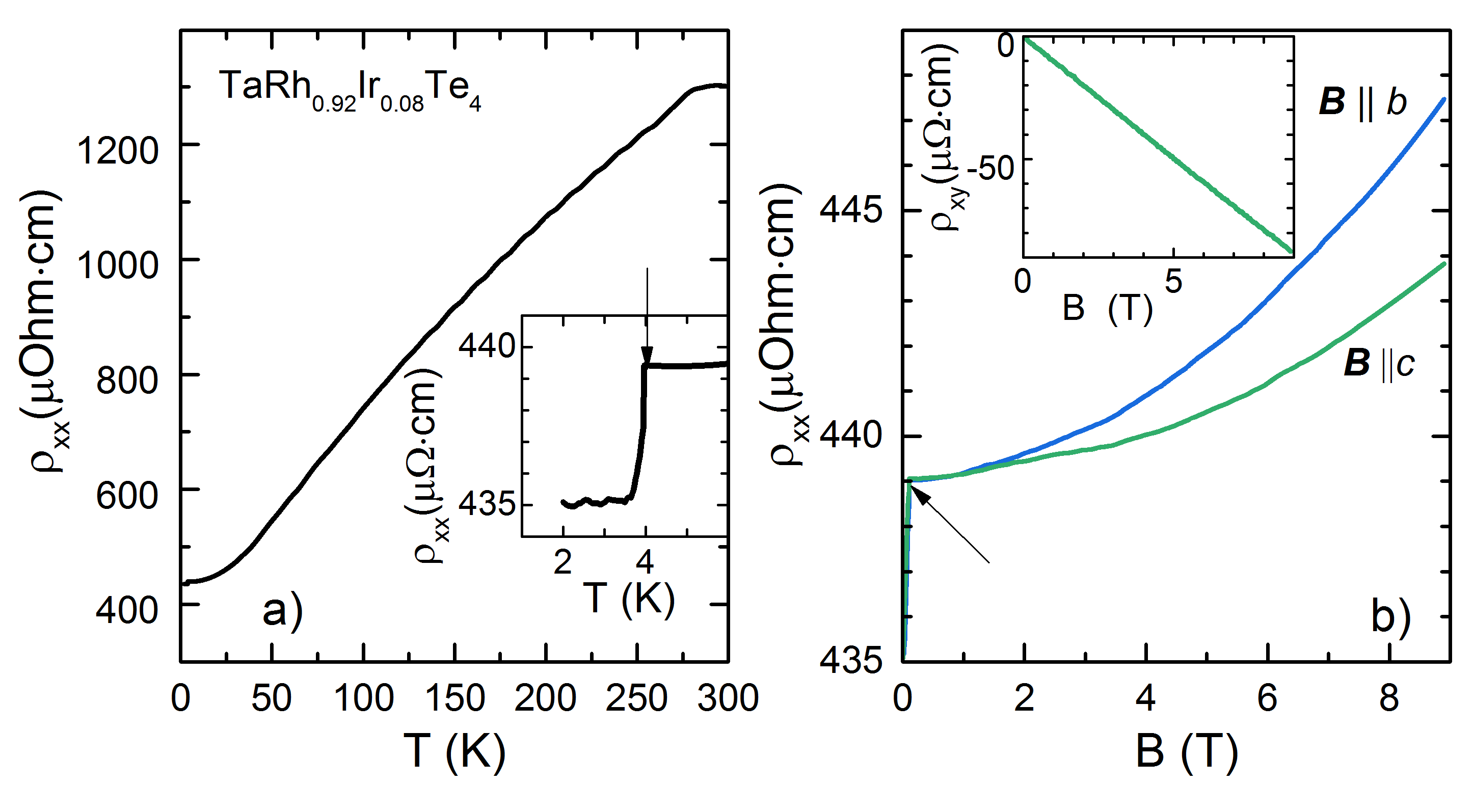}
\includegraphics[width=1.1\columnwidth{}]{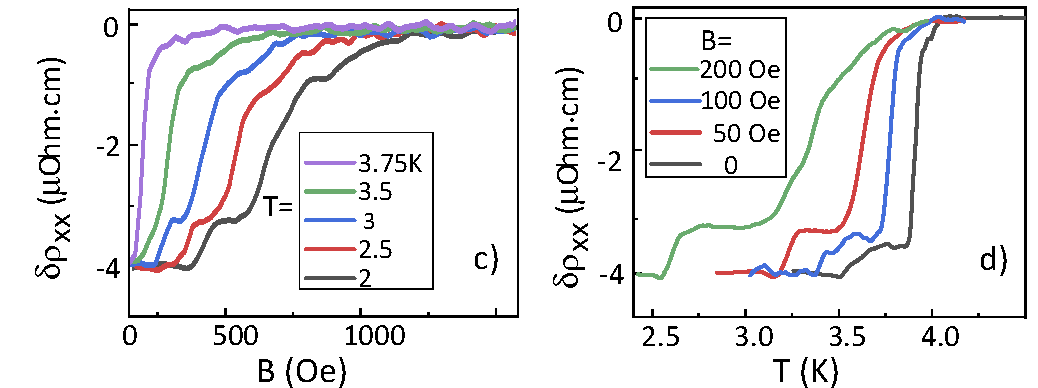}
\caption {(a) Temperature dependence of $\rho_{xx}$ for
  TaRh$_{\rm 0.92}$Ir$_{\rm 0.08}$Te$_4$.  Inset shows low temperature
  region of $\rho_{xx}(T)$.  (b) $\rho_{xx}$ versus magnetic field at
  two field orientations, measured at $T=2$K.  Inset: magnetic field
  dependence of the Hall resistivity $\rho_{xy}$. Arrows on panels (a)
  And (B) point at the onset of the SC transition; (c) low-field
  region of the $\rho_{xx}(H)$ dependencies measured at several fixed
  temperatures.  (d) low-temperature region of the $\rho_{xx}(T)$
  dependencies measured at several field values at $T=2$\,K.}
    \label{fig:RhIr008-SC}
\end{figure*}

We observed an indication of emerging superconductivity (SC) with
critical temperature $T_c = 4$\,K in \RhIr{} compound for $x=0.08$ and
in TaIr$_{1-x}$Ru$_x$Te$_4$ for $x=0.22$. As an example, we show in
Fig.~\ref{fig:RhIr008-SC} the data for the former compound.

The inset in Fig.~\ref{fig:RhIr008-SC} zooms in a sharp drop of
$\rho_{xx}(T)$ by 4$\,\mu$ Ohm$\cdot$cm both in zero field as
temperature decreases below $T\approx 4$\,K, and at $T=2$K as magnetic
field decreases below 0.1\,T. These features suggest that the drop of
$\rho_{xx}$ is related with the SC transition at $T_c\approx 4$\,K in
zero field.  Indeed, the transition temperature shifts to lower values
as magnetic field increases, and the critical magnetic field decreases
as temperature raises.

The zero-resistance state is not achieved, apparently because of a
very small fraction of the superconducting state.  In strong magnetic
field, MR shows weak classical quadratic dependence, almost isotropic
in $\theta$.  These slightly doped \RhIr{} and TaIr(Ru)Te$_4$ samples,
most likely, contain minor inclusions of several SC phases which
emerge in a narrow interval of doping on the compound phase diagram.
The interesting issue of potential superconductivity in these
compounds requires additional more detailed studies.

\FloatBarrier{}

\bibliography{TaTMTe4}

\begin{figure*}
  \includegraphics[width=0.55\textwidth]{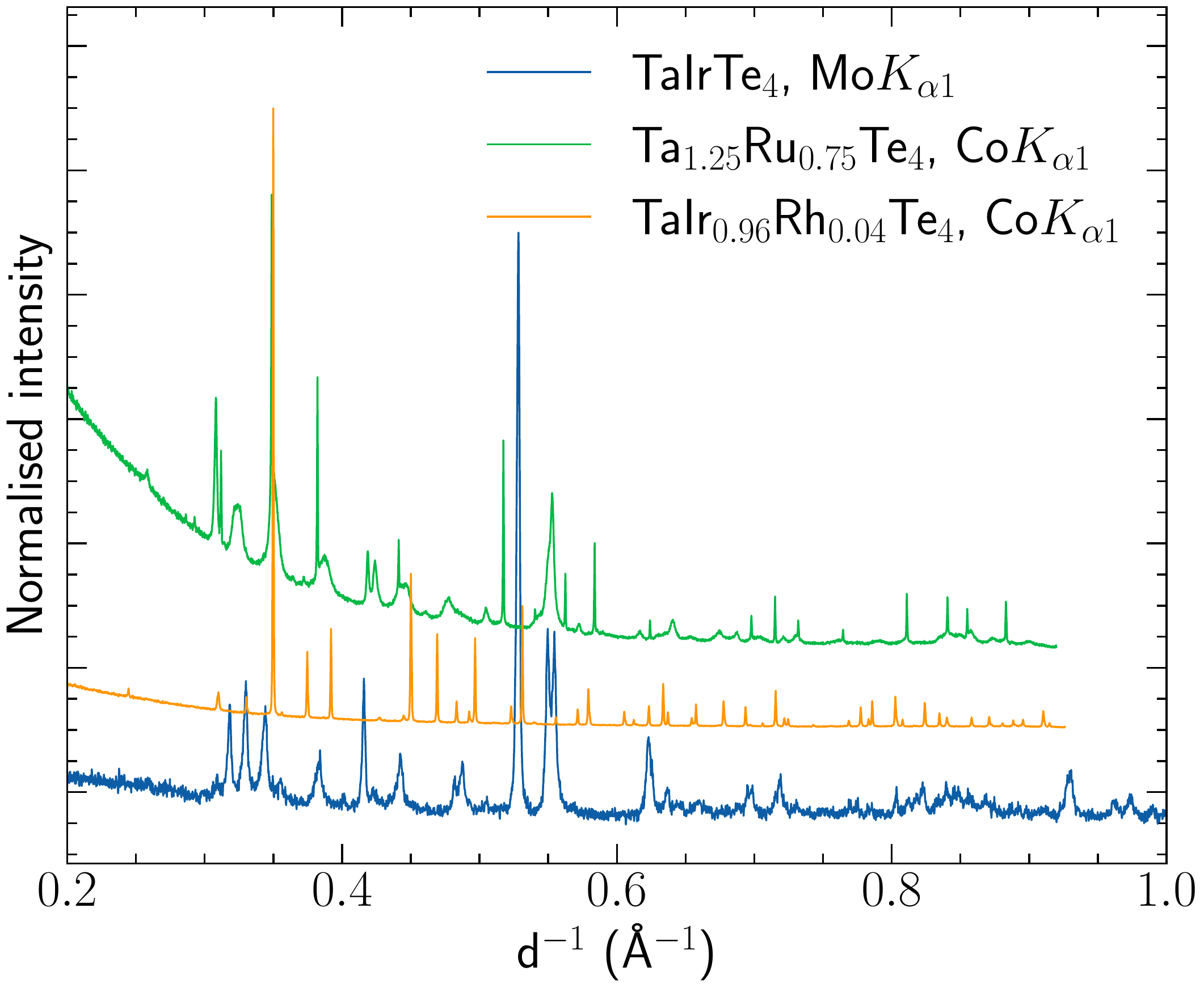}
  \caption{\label{pxrd-comp}Powder x-ray diffraction patterns for \Ir{}, \RuEDX{} and \IrRh{} (x = 0.04).}
\end{figure*}

\begin{table*}
  \caption{\label{tab:EDX:IrRh}
    comparison of nominal and composition measured in EDX for \IrRh{} crystals}
  \begin{tabularx}{\columnwidth}{Xl}
    \toprule
    Ir:Rh, nominal        & EDX composition              \\
    \midrule
    \ce{Ir_{0.9}Rh_{0.1}} & \ce{TaIr_{0.96}Rh_{0.04}Te4} \\
    \ce{Ir_{0.7}Rh_{0.3}} & \ce{TaIr_{0.82}Rh_{0.14}Te4} \\
    \ce{Ir_{0.3}Rh_{0.7}} & \ce{TaIr_{0.22}Rh_{0.78}Te4} \\
    \ce{Ir_{0.2}Rh_{0.8}} & \ce{TaIr_{0.08}Rh_{0.92}Te4} \\
    \bottomrule
  \end{tabularx}
\end{table*}

\begin{table*}
  \caption{\label{wyckoff}
    Refined atomic coordinates for \Rh{}}
  \begin{tabularx}{\columnwidth}{XXXrrrrrr}
    \toprule
    No & atom & label & x       & y       & z        & Occ.  & Site & Sym. \\
    \midrule
    1  & Ta   & Ta1   & 0.00000 & 0.05580 & -0.19200 & 1.000 & 2a   & m..  \\
    2  & Ta   & Ta2   & 0.00000 & 0.26530 & 0.29100  & 1.000 & 2a   & m..  \\
    3  & Rh   & Rh1   & 0.00000 & 0.52760 & -0.19800 & 1.000 & 2a   & m..  \\
    4  & Rh   & Rh2   & 0.00000 & 0.75430 & 0.33100  & 1.000 & 2a   & m..  \\
    5  & Te   & Te1   & 0.00000 & 0.05945 & 0.19049  & 1.000 & 2a   & m..  \\
    6  & Te   & Te2   & 0.00000 & 0.19454 & 0.62601  & 1.000 & 2a   & m..  \\
    7  & Te   & Te3   & 0.00000 & 0.34483 & -0.09190 & 1.000 & 2a   & m..  \\
    8  & Te   & Te4   & 0.00000 & 0.40783 & 0.44478  & 1.000 & 2a   & m..  \\
    9  & Te   & Te5   & 0.00000 & 0.57274 & 0.19981  & 1.000 & 2a   & m..  \\
    10 & Te   & Te6   & 0.00000 & 0.67725 & 0.65759  & 1.000 & 2a   & m..  \\
    11 & Te   & Te7   & 0.00000 & 0.85038 & -0.07740 & 1.000 & 2a   & m..  \\
    12 & Te   & Te8   & 0.00000 & 0.90438 & 0.46746  & 1.000 & 2a   & m..  \\
    \bottomrule
  \end{tabularx}
\end{table*}

\end{document}